# Compatibility and Accuracy Verification of CADmesh-Based Complex Geometry Modeling in Geant4

Weiyang Zhang [a], Shiwei Jing [a, *], Shengduo Liu [a], Jia Song [a], Sijia Zhou [a], Hailong Xu [a], Yue Sun [a], Zebin Li [a], Yuxuan Gu [a], Siqi Liu [a], Tian Zhang [a], Zhihua Gao [a], Guofeng Qu [a, *], Fuquan Jia [b]

[a] School of Physics, Northeast Normal University, Changchun 130024, Jilin, China

[b] Jilin Jianzhu University, Changchun 130118, China

*Corresponding author.

E-mail address: jingsw504@nenu.edu.cn (Shiwei Jing), quguofeng@scu.edu.cn (Guofeng Qu), 249404875@qq.com (Fuquan Jia).

**Abstract:** Geant4 Monte Carlo simulation relies on the Constructive Solid Geometry (CSG) method for complex geometric modeling. This method has low efficiency and a high application threshold. Importing triangular facet formats such as STL/OBJ via CADmesh is a promising alternative, but systematic evaluations of format compatibility, geometric accuracy, and physical simulation deviations are lacking. Construct open-source experimental environment based on Geant4 11.0, CADmesh 1.3.0 and FreeCAD 1.0. We design high and low precision gradient test cases using simple geometric bodies and complex engineering models, and systematically evaluate the import success rate, facet loss rate, volume error, and particle transport dose deviation for STL and OBJ formats. The results show a 100% import success rate for both formats; the volume error rate is ≤0.018% for high-precision models and ≤0.288% for low-precision models. The two formats share the same vertex facet data structure. This study designs a general adaptive interface. The interface reduces the number of parsing code lines by about 70% and maintains geometric accuracy. Furthermore, the tetrahedral mesh loading takes 3.1 times longer than tessellated solids, but the simulation time can be reduced from 15194.3 s to 77.28 s.

# 1. Introduction

Geant4 is an open-source Monte Carlo particle transport simulation toolkit developed by CERN, widely used in nuclear physics, radiation protection, and other fields [1]. Its core geometric modeling relies on constructive solid geometry (CSG), where users manually write C++ code to define basic geometric shapes and assembly relationships. This approach not only requires proficient programming skills but also leads to long modeling cycles [2] and introduces human errors when dealing with complex engineering models.

Researchers have proposed various technical paths for importing CAD models into Geant4. The GDML method requires two-step conversion, which may cause loss of geometric information [3]. The MCAM program developed by the FDS team performs well in scenarios such as fusion reactors, but its optimization for general mesh formats is limited. The commercial software FastRAD has licensing restrictions [4]. CADmesh is an open source C++ library that directly imports triangular facet mesh files such as STL and OBJ into Geant4, and has been applied in dose calculation and other areas [5]. However, existing studies have not systematically evaluated the compatibility of STL and OBJ formats, and quantitative analyses of facet loss, volume error, and dose deviation after import are scarce.

The main contributions of this paper are: building a standardized experimental environment, designing high- and low-precision gradient test cases, systematically evaluating the compatibility performance of STL/OBJ formats; quantitatively analyzing geometric accuracy and physical accuracy; designing a universal adapter interface based on the common data structure of the two formats; and providing accuracy optimization strategies for practical applications.

# 2. Materials and Methods

## 2.1 Experimental setup

The experiments use stable and open-source software versions. Geant4 version 11.0, the current mainstream stable release, is chosen for its optimized parsing efficiency of the G4 TessellatedSolid class and full compatibility with the latest CADmesh interface [6]. CADmesh version 1.3.0 supports STL/OBJ format parsing, and its C++ interface is fully compatible with Geant4 11.0 [7]. FreeCAD 1.0 is used for parametric precision modeling to ensure geometric dimensional accuracy. The environment is validated by importing a standard geometric body. Taking a sphere of radius 1 m as an example, exporting STL (ASCII encoding, tolerance 0.01 mm) and importing into Geant4 via CADmesh yields the same number of facets (12288) as during export, with vertex coordinate deviations less than $1\times10^{-6}$ m, meeting accuracy requirements.

## 2.2 Test models

Comprehensively cover simple and complex scenarios and ensure the generality and specificity of compatibility analysis, two types of test geometries are designed: simple standard geometric bodies and complex models. All models are generated via parametric modeling in FreeCAD [8]. The specific parameters and design justifications are as follows:

Sphere: radius 0.5 m, no hollow structure. The facet tolerance during FreeCAD export is set to two levels: 0.01 mm (high precision) and 1 mm (low precision), corresponding to STL and OBJ formats, yielding a total of 4 test files for each shape;

Cylinder: base radius 0.3 m, height 1 m, axis along Z. Facet tolerances: 0.01 mm (high precision) and 1 mm (low precision), corresponding to STL and OBJ formats, giving 4 test files.Simplified detector housing model: a “cylinder+hemisphere” composite structure,with outer diameter 1 m, height 2 m, wall thickness 5 cm, top hemisphere radius 0.5 m, and three evenly distributed cylindrical hollows (radius 0.1 m, depth 1.5 m) inside. Constructed using FreeCAD’s assembly workbench, exported in STL and OBJ formats (including material: housing aluminum, hollow regions vacuum), with two files each (facet tolerances 0.05 mm/0.1 mm). The design of the standard geometric bodies follows Geant4’s geometric accuracy requirements [9]; the dimensions cover typical scales of detector components and experimental apparatus structures commonly encountered in nuclear physics simulations. The facet tolerance gradient is used to subsequently analyze the impact of precision on compatibility [10].

## 2.3 Import and preprocessing workflow

Systematic tests are carried out according to the relevant workflow. Verification schemes are designed for the core characteristics of STL and OBJ formats to quantitatively evaluate the adaptation effect. The test flowchart is shown in Fig. 1.

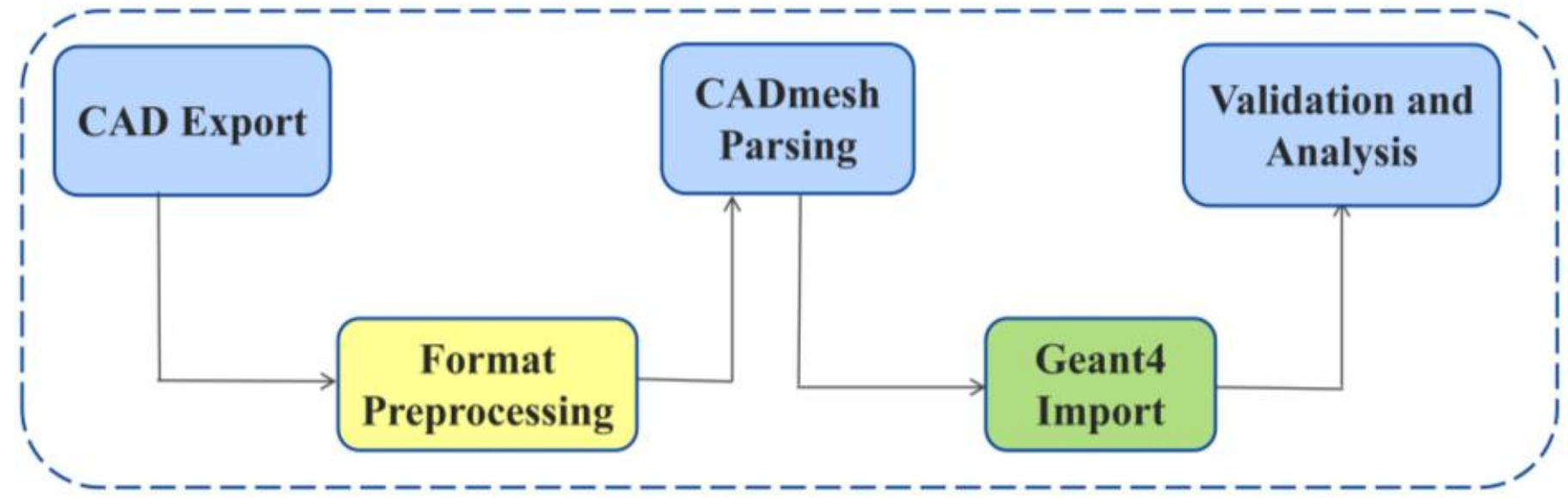

**Fig. 1** Flowchart of STL and OBJ format import testing.

Selecting three simple geometric bodies (cube, frustum, sphere) and one complex geometric body (a neutron activation model with 18,544 vertices and 37,984 facets) are selected as test cases, loaded as tessellated solids and tetrahedral meshes, respectively, as shown in Fig. 2.

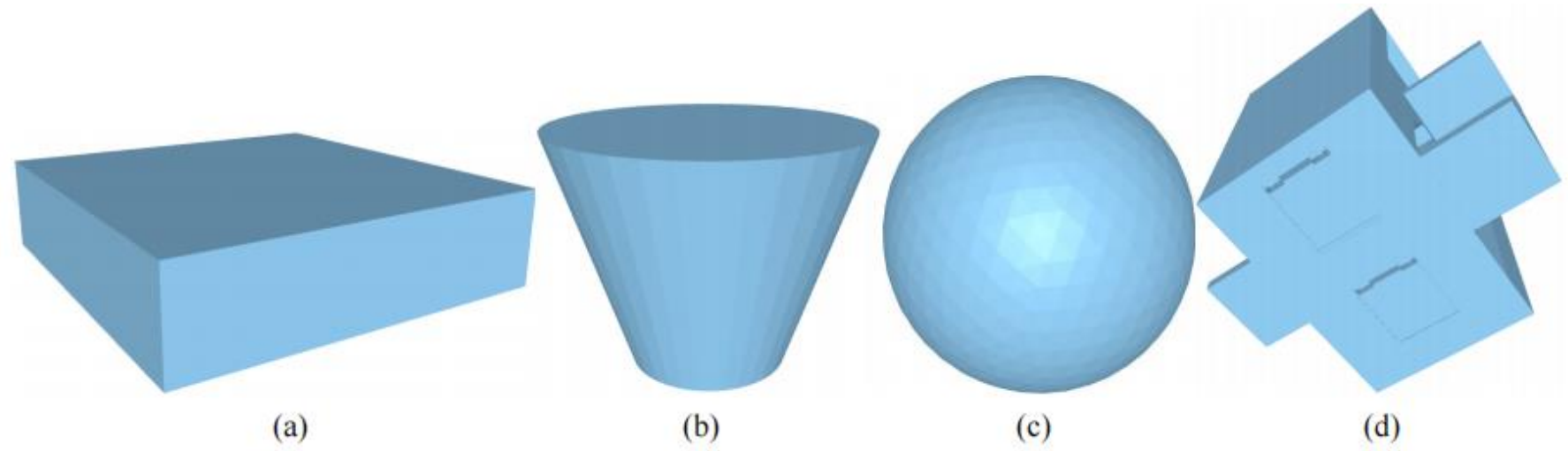

**Fig. 2** Diagrams of simple and complex geometries: (a) cube; (b) frustum; (c) sphere; (d) neutron activation model.

Using the C++ interface of CADmesh1.3.0, complete CAD model export, format preprocessing, CADmesh parsing and Geant4 integrated import, and adjust preprocessing and parsing methods based on import characteristics of simple and complex models. Use MeshLab 2023.12 to handle common problems of files exported from FreeCAD such as scattered surface facets of simple models and connecting holes between components of complex models. Process simple models with cleaning repair and duplicate vertex removal to eliminate small gaps on surfaces, and conduct three times boundary-retained Laplacian smoothing to balance surface facet distribution. Deal with complex models by repairing non-manifold edges to fill holes no larger than 0.01 meters, removing overlapping faces and finishing mesh reconstruction and simplification. Directly cover original STL files and update corresponding .mtl files for OBJ files to keep normal material matching. Practical test data shows the spherical mesh tolerance is 0.01mm, simplified facet quantity falls from 150000 to 70000, import time shortens from 32 seconds to 14 seconds, and geometric volume error rate changes from 1.15% to 1.32%.

## 2.4 Design of a universal adapter interface

The compatibility test results of STL and OBJ formats and the common underlying data structure of the two formats, a universal data interface is designed to achieve unified parsing and Geant4 adaptation for both formats, and the exploration process and results are presented objectively.

First, the core commonality between STL and OBJ formats is the vertex facet geometric description logic; the geometric core data of both are consistent, including vertex coordinates and facet indices. The triangular facets of STL are fully consistent with triangulated facets of OBJ. Second, the parsing processes are similar: both require reading vertex data first, then facet indices, and finally constructing topological relationships; the adaptation target for Geant4 is the same, i.e., conversion into geometric entities of the `G4TessellatedSolid` class, with identical encapsulation logic [11]. The core design of the universal interface is to “extract common parsing logic and shield format differences,” eliminating the need to write separate parsing code for each format and improving adaptation efficiency.

The universal interface is developed in C++, based on the underlying API of CADmesh, and follows a four-stage process:format recognition – common parsing – difference handling Geant4 encapsulation. First, the format recognition module automatically identifies the format via file extension (`.stl`/`.obj`) and calls the corresponding format validation function to verify whether the file header conforms to STL/OBJ standards. Second, the common parsing module uniformly reads vertex coordinates and facet indices and stores them in a common data structure. The difference handling module processes format-specific information separately: STL requires no additional handling; OBJ reads material information and maps it to the Geant4 material library [12]. Finally, the Geant4 encapsulation module encapsulates the vertex and facet data from Common Geometry Data into a G4TessellatedSolid, uniformly associates logical volumes and physical volumes, and completes the import.

## 2.5 Accuracy evaluation

A comparative experiment is designed between FreeCAD modeling import and native Geant4 coding modeling, covering the target models: sphere, cube, frustum, and neutron activation model. The deviation is quantified from both geometric and physical accuracy dimensions to verify the feasibility of FreeCAD based modeling. Native code is written for each model with parameters exactly matching those in FreeCAD modeling (dimensions, materials, positions) as the accuracy benchmark [13].

Geometric accuracy metrics: Key vertices (e.g., sphere poles, frustum top and bottom circle centers, cube center) are selected; coordinates are extracted using Para View; Euclidean distance [14] deviation is calculated as Eq. (1).The maximum and average values are recorded. Additionally, the volume error rate is calculated as Eq. (2).

$$\triangle d = \sqrt{(\mathcal{X}_{\text{FreeCAD}} - \mathcal{X}_{\text{Native}})^2 + (\mathcal{Y}_{\text{FreeCAD}} - \mathcal{Y}_{\text{Native}})^2 + (\mathcal{Z}_{\text{FreeCAD}} - \mathcal{Z}_{\text{Native}})^2} \quad (1)$$

$$\triangle \mathcal{V} = \frac{|\mathcal{V}_{FreeCAD} - \mathcal{V}_{Native}|}{\mathcal{V}_{Native}} \times 100\% \quad (2)$$

For the complex model, dose grids are placed at key cross-sections z = 1.0 m and z = 0.25 m with grid cell size 0.005 m × 0.005 m. The relative deviation of dose distribution [15] is used as the evaluation metric, as shown in Eq. (3).

$$\triangle \mathcal{D} = \frac{\left|\mathcal{D}_{FreeCAD(x,y,z)} - \mathcal{D}_{Native(x,y,z)}\right|}{\mathcal{D}_{Native}} \times 100\% \quad (3)$$

# 3. Experiments and discussion

## 3.1 Compatibility evaluation

The stored geometric information of all test geometries number of triangular facets, number of vertices, volume is highly consistent with the faceting processing logic of CADmesh. The geometric information integrity and parsing stability are verified following the test flowchart, with custom code to generate facet count, vertex count, and volume.

(1) Qualitative analysis

The Geant4 visualization interface shows: after importing STL and OBJ files with high precision (tolerance ≤0.05 mm). At the same time the required number of triangular facets is large and each facet area is small. These tiny facets, when assembled, visually approximate a smooth continuous spherical surface, with almost no visible "faceted edges". The curved surfaces of spheres and cylinders transition smoothly; for the complex engineering model, the hollow structures and component junctions show no missing or overlapping facets, and the structure is consistent with the original CAD model. The constructed models are shown in Fig. 3.

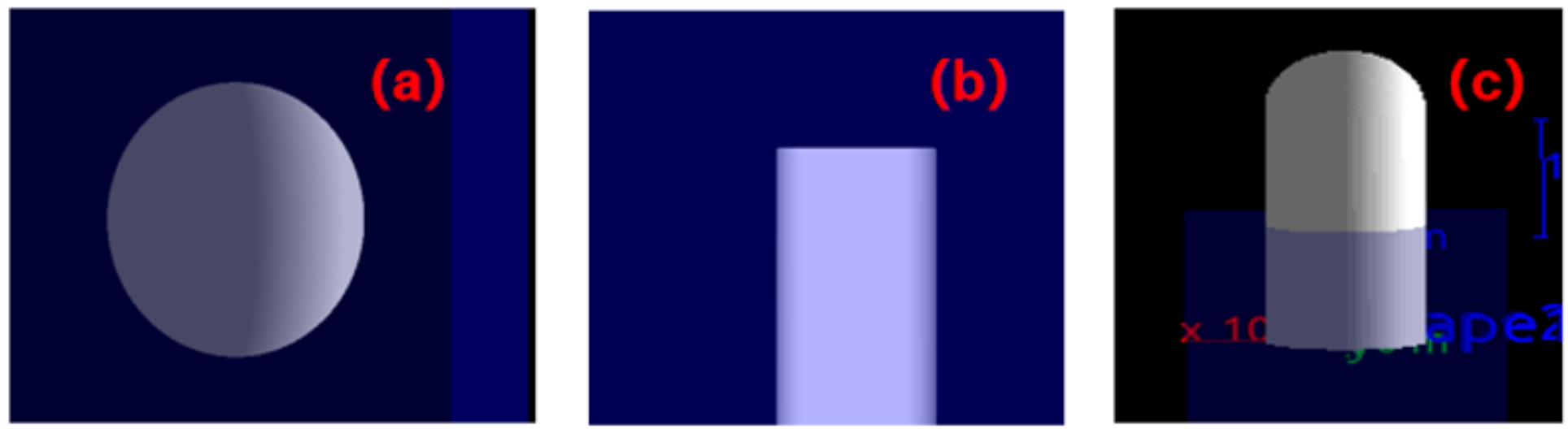

**Fig. 3** High-precision constructed models: (a) sphere; (b) cylinder; (c) detector housing.

After importing STL and OBJ files with low precision (tolerance = 1 mm), the number of triangular facets is small and each facet area is large, resulting in an obvious "faceted" appearance on the spherical surface and prominent polygonal edges. The curved surfaces exhibit pronounced polygonal distortion, as shown in Fig. 4.

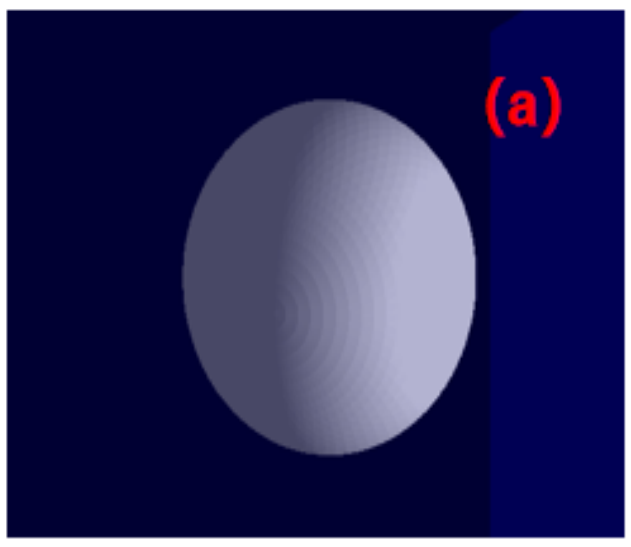

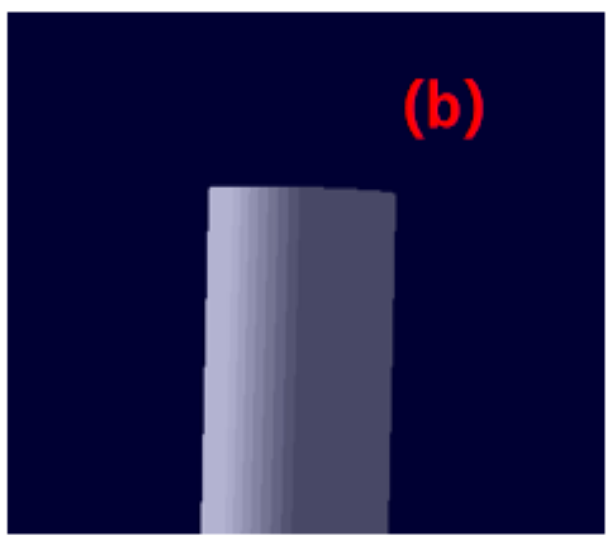

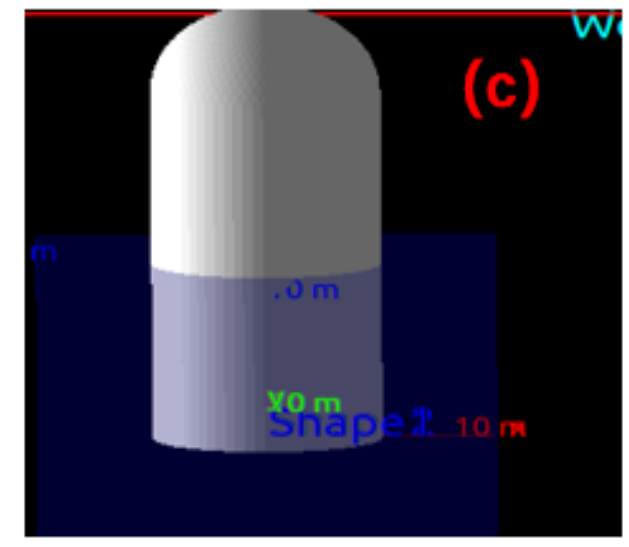

**Fig. 4** Low-precision constructed models: (a) sphere; (b) cylinder; (c) detector housing.

(2) Quantitative analysis

Compile Geant4 geometry import program code, generate visual geometry, output facet count, vertex count and volume of imported geometry, calculate volume error rate and successful import rate. Theoretical volumes of the sphere and cylinder are calculated using Formula (4) and Formula (5) respectively.

$$V_{\text{sphere}} = \frac{4}{3}\pi r^3 \tag{4}$$

$$V_{cylinder} = \pi r^2 h \tag{5}$$

On this basis, calculate volume error rate according to volume deviation before and after import using Formula (6), and calculate successful import rate according to file import results using Formula (7).

$$Volume\ error\ rate = \frac{|Measured\ value - True\ value|}{True\ value} \times 100\% \tag{6}$$

$$Successful\ import\ rate = \frac{Number\ of\ successfully\ imported\ files}{Total\ number\ of\ test\ files} \times 100\% \tag{7}$$

After import, the geometric data of simple and complex models are extracted and compared with the original CAD models. The verification results of geometric information integrity are shown in Table 1.

**Table 1** Verification results of geometric information integrity.

| Model Type | Precision Level | Original Number of Faces | Number of Faces After Import | Volume Error Rate(%) | Import Success Rate（%） |
|---|---|---|---|---|---|
| Sphere | High | 100518 | 100518 | 0.015 | 100 |
| Sphere | Low | 5006 | 5003 | 0.288 | 100 |
| Cylinder | High | 2176 | 2176 | 0.002 | 100 |
| Cylinder | Low | 216 | 214 | 0.217 | 100 |
| Detector Housing | High | 436034 | 436013 | 0.018 | 100 |
| Detector Housing | Low | 4696 | 4696 | 0.235 | 100 |

The STL format has excellent basic compatibility with Geant4: all test files were successfully imported. High-precision files have a volume error rate ≤0.015% compared to the original model, meeting the geometric accuracy requirements of Geant4 particle transport simulations [16]; low-precision files (tolerance ≥1 mm) have a facet loss rate ≤0.67% and a volume error rate ≤2.13%.

Sort out common problems in STL and OBJ format import process, classify and analyze by phenomenon, cause and case, propose solutions such as preprocessing and parameter adjustment, provide reference for practical application.Facet loss missing facets appear in local regions of the imported model. This is particularly evident in low-precision STL files and complex models. The causes are: excessively large facet tolerance during CAD export, leading to undersampling when curved surfaces are discretized into triangular facets, creating gaps between facets; slight displacement during assembly of complex models, causing facet overlap at junctions; STL format lacks topological description, facets are defined only by vertex coordinates, and when tolerance is too large, they cannot be merged automatically, resulting in gaps.

## 3.2 Performance of the universal adapter interface

Verifying the effectiveness and generality of the universal adapter interface, a comparative experiment is designed. Two groups are set: the control group uses the standard CADmesh calling method, requiring separate parsing code for STL and OBJ formats; the experimental group uses the proposed universal interface to parse both formats uniformly. Three typical models sphere high precision, cylinder high precision, detector housing high precision are selected. Evaluation is performed from two dimensions: functional completeness and processing efficiency. Each model is tested 10 times and the average is taken.

(1) Functional completeness comparison

The differences between the two methods are compared in terms of code complexity, format support, and extensibility, as shown in Table 2.

**Table 2** Functional analysis of standard CADmesh call vs. universal interface.

| Comparison Aspect | Standard CADmesh | Universal Interface |
|---|---|---|
| Lines of Code | 15 lines required for STL, 25 lines required for OBJ | 10 lines required for unified invocation |
| Format Recognition | Manual judgment of file extensions with separate invocations | Automatic recognition, no manual intervention required |
| Code Maintenance | Independent maintenance | Single unified maintenance path |
| Extensibility | Requires rewriting parsing modules for new formats | Extensible to new formats based on common structure |

The universal interface is significantly superior to standard CADmesh calls in terms of functional completeness: code volume is reduced by approximately 70%, manual format judgment is eliminated, material information is automatically mapped, and development and maintenance costs are greatly reduced. Moreover, based on the common vertex facet data structure, the universal interface can be extended to new formats by simply extending the format recognition module while reusing the parsing logic, demonstrating good extensibility.

(2) Processing efficiency comparison

The processing efficiency is compared from three aspects: parsing time, memory usage, and geometric fidelity, as shown in Table 3.

**Table 3** Performance analysis of standard CADmesh call vs. universal interface.

| Model | Metric | Standard CADmesh Invocation | Universal Interface |
|---|---|---|---|
| Sphere (High Precision) | Parsing Time (s) | 1.21 | 1.18 |
| | Memory Usage (MB) | 245 | 242 |
| | Volume Error Rate (%) | 0.015 | 0.015 |
| Cylinder (High Precision) | Parsing Time (s) | 0.78 | 0.76 |
| | Memory Usage (MB) | 156 | 154 |
| | Volume Error Rate (%) | 0.002 | 0.002 |
| Detector Housing (High Precision) | Parsing Time (s) | 4.56 | 4.51 |
| | Memory Usage (MB) | 892 | 886 |
| | Volume Error Rate (%) | 0.018 | 0.018 |

The universal interface outperforms the standard CADmesh call slightly: parsing time is reduced by an average of 1.5%–2.6%, and memory usage is reduced by an average of 0.7%–1.3%. The efficiency improvement is mainly due to the unified data structure, which reduces redundant overhead from format recognition and intermediate conversion steps. In terms of geometric fidelity, the volume error rates of the two methods are identical, indicating that the universal interface does not introduce additional geometric information loss during encapsulation, and geometric accuracy is preserved. By writing and testing the interface code, the universal interface successfully achieves unified import of STL ASCII and OBJ formats without distinguishing format types; the interface call is concise, requiring only one line of code. After importing OBJ and STL files through the universal interface, the geometric fidelity is consistent with separate imports, with no additional facet or volume loss.

Therefore, the exploration of the universal interface has achieved universal adaptation in some scenarios, verifying the feasibility of designing a universal interface based on common data structures. However, given the diversity and complexity of models, complete universality may not yet be achieved; further improvements in model adaptation and efficiency optimization are needed.

## 3.3 Accuracy verification

After importing the FreeCAD-modeled geometries, the volume error rates of the sphere and frustum are both ≤0.15%, reaching the excellent level, verifying the precision of FreeCAD parametric modeling. The deviation mainly originates from facet discretization: STL and OBJ are facet-based models, whereas native code uses analytical geometry; however, the deviation magnitude is much lower than Geant4 particle transport uncertainties. The average dose deviation is ≤1.08% and the maximum deviation is ≤2.93%, meeting the excellent requirement. The deviation is attributed to the curved surface approximation in facet models, causing slight differences in particle path calculation, but the effect on overall dose distribution trends is negligible.

For the complex model, the vertex deviation and volume error rate are higher than for simple models, with volume error rate ≤1.39%, still in the good range. The deviation mainly comes from facet errors at multi component junctions and residual topology optimization after Boolean operations. The average dose deviation for the detector housing with STL import is 2.15%, and with OBJ import it is 2.63%, both acceptable. The accuracy of STL format is generally better than that of OBJ format. The reason is that OBJ requires additional parsing of material information; FreeCAD may introduce material association deviations during export, and facetization rules are more complex, supporting quadrilateral facets that require automatic triangulation by CADmesh, increasing error.

## 3.4 Accuracy optimization attempt

Given that the physical accuracy of the complex detector model is close to the acceptable threshold, two optimization strategies are designed: facet tolerance adjustment and model detail simplification. Comparative experiments are carried out to validate the effects, and the optimization logic and limitations are objectively analyzed. First, the facet tolerance during FreeCAD modeling is reduced to increase facet density, improving geometric fidelity of complex structures. For complex models, a combined strategy of “facet tolerance adjustment + model detail simplification” is recommended. For example, use tolerances of 0.02 mm for outer parts and 0.01 mm for internal parts (STL format), and remove irrelevant details such as small chamfers and surface textures. This strategy is suitable for scenarios where the key regions for particle transport require high accuracy and computational resources are sufficient. For ultra-complex models with facet counts in the millions, this strategy may lead to unacceptably low computational efficiency, and further optimization using Geant4 simplified geometry tool `G4SubtractionSolid` or other options should be considered.

### 3.5 Simulation performance comparison

OpenGL viewer is used to qualitatively inspect the geometric structures, checking for obvious structural damage, and whether tessellated solids and tetrahedral meshes are correctly positioned within their mother volumes. The G4TessellatedSolid::DumpInfo() method is called to store tessellated solid and tetrahedral data, extract facet and vertex coordinate information, and compare point-by-point with the original CAD model. Built in geometry overlap test is used to ensure adjacent volumes touch only at boundaries without overlap. Finally, Geant4 pseudo particles are emitted from three orthogonal directions into the geometry, with tracking detail set to 1 mm, so that the output includes the unique name of each volume navigated as traverse the user detector geometry. The output is searched to verify that both tessellated solids and tetrahedral meshes are navigable and are included as volumes in the Geant4 user detector geometry.

In the qualitative inspection, the OpenGL viewer shows no geometric damage; the model morphology is complete and correctly positioned within the mother volume. The tessellated solids of the four test geometries are shown in Fig. 5.

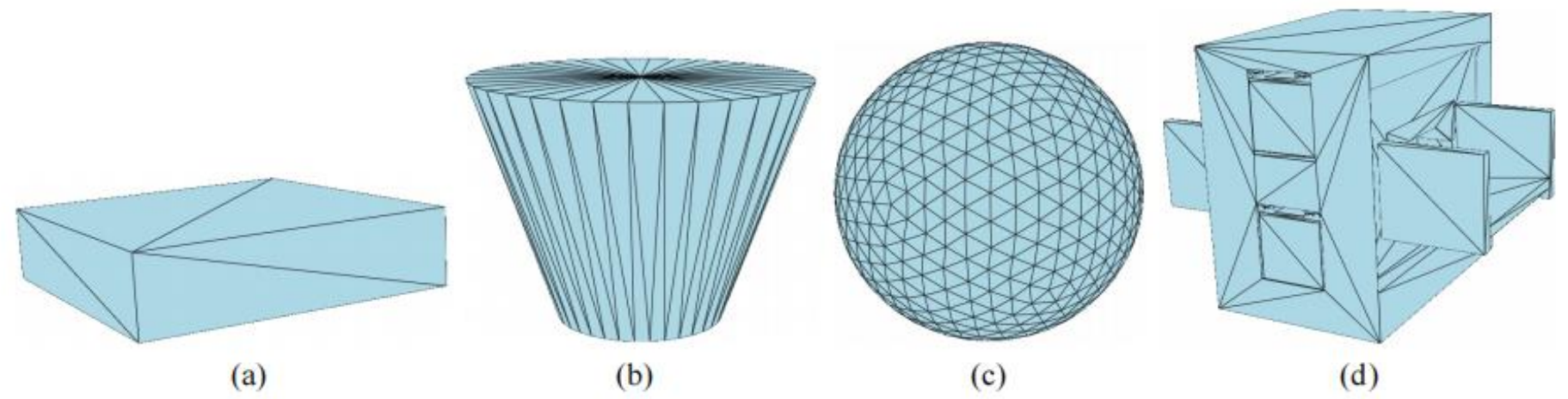

**Fig. 5** Tessellated solids of the four test geometries: (a) cube; (b) frustum; (c) sphere; (d) neutron activation model.

After analysis, the tetrahedral meshes of the four test geometries are shown in Fig. 6.

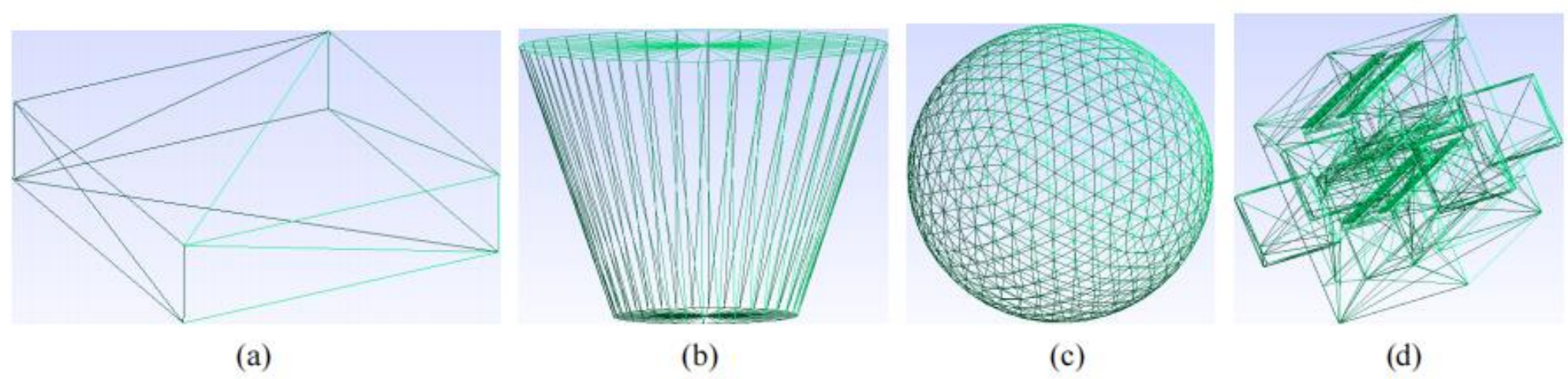

**Fig. 6** Tetrahedral meshes of the four test geometries: (a) cube; (b) frustum; (c) sphere; (d) neutron activation model.

Geant4's geometry overlap test shows that all adjacent volumes touch only at boundaries without overlap. The results reported by `G4TessellatedSolid::DumpInfo()` are consistent with the original CAD model data; the tessellated solids and tetrahedral meshes have no differences in facet or vertex coordinates from the original CAD model. Emitting geantinos from three orthogonal directions shows that both tessellated solids and tetrahedral meshes are correctly loaded into Geant4 and can be used for subsequent operations.

Verifying simulation performance, a neutron activation model made of G4WATER is placed in a G4AIR cubic world of side length 4 m. The geometry is loaded as either a native Geant4 tessellated solid or a tetrahedral mesh converted by the CADmesh system. Simulations are run with 10 particles, repeated 10 times to reduce statistical error. A schematic of the simulation is shown in Fig. 7.

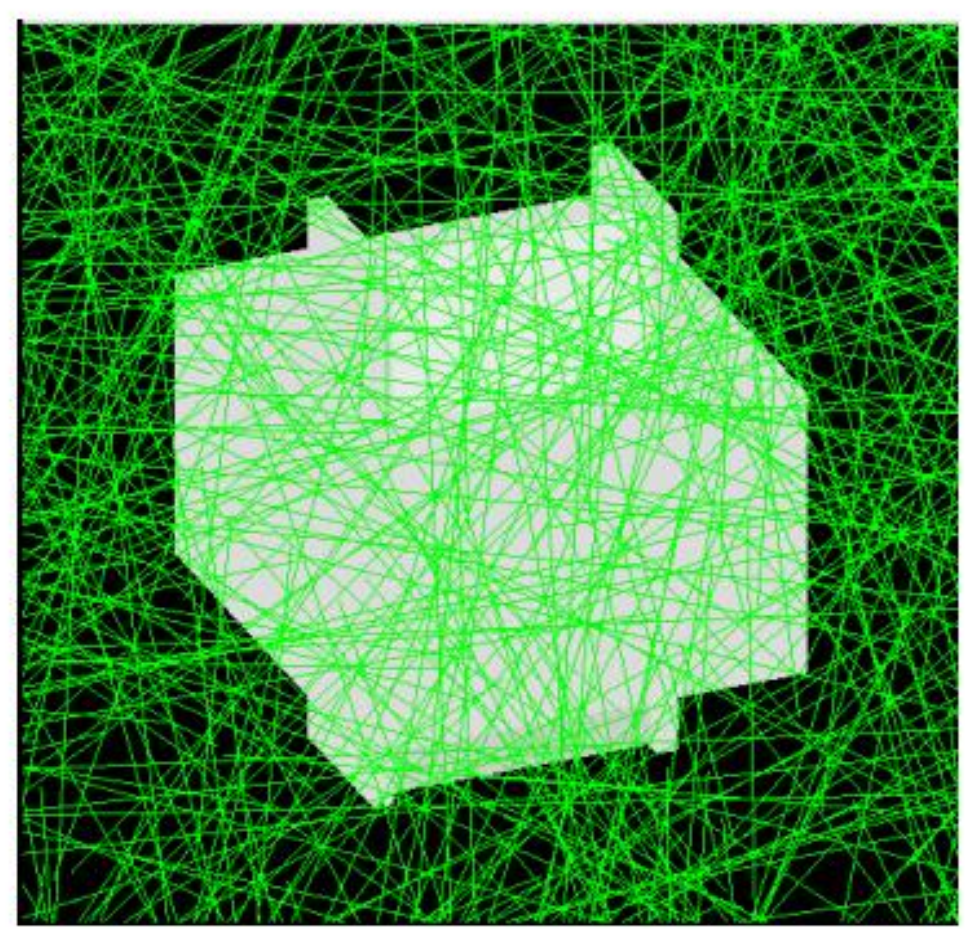

**Fig. 7** Schematic of particle bombardment simulation experiment.

By recording the geometry loading time and total simulation time, the performance difference between tessellated solids and tetrahedral meshes is compared. The loading time for the tessellated solid is the total time to add all facets to `G4TessellatedSolid` and position them within the user detector geometry. The loading time for the tetrahedral mesh is the cumulative time to initialize each tetrahedral unit as a `G4Tet` entity, add it to the assembly volume, and complete assembly in the user detector geometry. Table 4 shows the geometry loading time and simulation time for the neutron activation model.

**Table 4** Loading time and simulation time for the neutron activation model.

| Time（s） | Embedded Solid (CSG) | Tetrahedral Mesh |
| --- | --- | --- |
| Loading Time | 91.23 | 282.53 |
| Simulation Time | 15194.36 | 77.28 |

The experimental results show that the loading time of the tetrahedral mesh is 282.53 s, 3.1 times that of the tessellated solid (91.23 s). This is because the tetrahedral mesh needs to handle many more geometric primitives, though both have complexity. However, in terms of simulation efficiency, the tetrahedral mesh achieves a simulation time of only 77.28 s, which is more than two orders of magnitude faster than the tessellated solid (15194.36 s). From a practical perspective, although the tetrahedral mesh takes longer to load, the loading process is negligible compared to simulation times which are often on the order of hours. Therefore, the difference in loading time is negligible, and tetrahedral meshes offer significant advantages for large-scale simulation scenarios.

## 4. Conclusion

Evaluate compatibility and accuracy of STL/OBJ format import into Geant4 via CADmesh, results show 100% import success rate for both formats, volume error rate ≤ 0.018% for high-precision files, facetization tolerance as core factor affecting parsing accuracy, general adaptation interface designed based on common vertex-facet data structure reduces code volume by about 70% while maintaining accuracy, simulation performance comparison shows tetrahedral mesh simulation efficiency much higher than tessellated solids, this study provides verification method and adaptation technical path for CAD model import into Geant4, future work focuses on expanding support for more mesh formats, developing automatic defect repair algorithms, and packaging visual import tool based on QT framework.

**CRediT authorship contribution statement**

**Weiyang Zhang:** Conceptualization, Methodology, Software, Data curation, Validation, Investigation, Formal analysis, Writing-original draft. **Shengduo Liu:** Software. **Jia Song:** Resources. **Sijia Zhou:** Formal analysis. **Hailong Xu:** Software, Visualization. **Yue Sun:** Software, Visualization. **Zebin Li:** Formal Analysis. **Yuxuan Gu:** Visualization. **Siqi Liu:** Formal analysis. **Tian Zhang:** Formal analysis. **Zhihua Gao:** Formal analysis. **Shiwei Jing:** Resources, Funding acquisition, Writing-review & editing, Conceptualization, Project administration. **Guofeng Qu:** Funding acquisition, Resources. **Fuquan Jia:** Formal analysis, Writing-review & editing.

**Acknowledgements**

This research was supported by the Education Department of Jilin Province of China (Grant No. JJKH20250303B